\documentclass[prd,aps,preprint,nofootinbib,superscriptaddress]{revtex4}
\usepackage{amsmath}
\usepackage{epsfig}
\usepackage{graphics}
\pdfoutput=1

\def\lsim{\raise0.3ex\hbox{$\;<$\kern-0.75em\raise-1.1ex\hbox{$\sim\;$}}}
\begin{document}

%\hfill{NPAC-08-24}

\title{Direct and Indirect Singlet Scalar Dark Matter Detection in the Lepton-Specific two-Higgs-doublet Model}

\author{M.~S.~Boucenna}
\email{boucenna@ific.uv.es}
\affiliation{AHEP Group, Institut de F\'isica Corpuscular C.S.I.C./Universitat de Val\`encia\\ Edificio Institutos de Paterna, Apt 22085, E-46071 Valencia, Spain}
\affiliation{Department of Physics, University of California, 1156 High St., Santa Cruz, CA 95064, USA}
\author{S.~Profumo}
\email{profumo@scipp.ucsc.edu} \affiliation{Department of Physics, University of California, 1156 High St., Santa Cruz, CA 95064, USA}\affiliation{Santa Cruz Institute for Particle Physics, Santa Cruz, CA 95064, USA} 

\date{\today}

\begin{abstract}
\noindent  A recent study of gamma-ray data from the Galactic Center motivates the investigation of light ($\sim7-10$ GeV) particle dark matter models featuring tau lepton pairs as dominant annihilation final state. The Lepton-Specific two-Higgs-doublet Model (2HDM-L) provides a natural framework where light, singlet scalar dark matter can pair-annihilate dominantly into tau leptons. We calculate the nucleon-dark matter cross section for singlet scalar dark matter within the 2HDM-L framework, and compare with recent results from direct detection experiments. We study how direct dark matter searches can be used to constrain the dark matter interpretation of gamma ray observations, for different dominant annihilation final states. We show that models exist with the correct thermal relic abundance that could fit the claimed gamma-ray excess from the Galactic Center region and have direct detection cross sections of the order of what needed to interpret recent anomalous events reported by direct detection experiments.
\end{abstract}

\maketitle

%\noindent\hrulefill
%
%\tableofcontents
%
%\noindent\hrulefill
%\newpage
%%%%%%%%%%%%%%%%%%%%%%%%%%%%%%%%%%%%%%%%%%%%%%%%%%%%%%%%%%%%%%%%%%%%%%%%%%%%%%%%%%%%%%%%%%%%%%%%%%%%%%%%%
%%%%%%%%%%%%%%%%%%%%%%%%%%%%%%%%%%%%%%%%%%%%%%%%%%%%%%%%%%%%%%%%%%%%%%%%%%%%%%%%%%%%%%%%%%%%%%%%%%%%%%%%%
\section{Introduction, Motivation and Outline}
%
%\noindent {\large\bf This section by SP} 

These are exciting times in the search for the particle nature of dark matter. A remarkable convergence of experimental capabilities is connecting direct detection campaigns with indirect searches and with the collider program (see e.g. Ref.~\cite{Profumo:2011zj}). Tantalizing signatures, and a few controversial claims, have focused some attention on a specific class of weakly interacting massive particle (WIMP) dark matter candidates, with relatively light masses (roughly, in the range between 7 and 10 GeV). Here, we briefly review these experimental data, and we give a detailed proof of existence of a particle setup (originally proposed in Ref.~\cite{Logan:2010nw}) that can successfully explain them, while at the same time having a thermal relic density matching the observed dark matter density in the universe. 

%cogent seems real, with modulation and it can be perhaps made compatible with null results from CDMS, Xenon, perhaps even DAMA
After reporting an irreducible excess of low-energy (below 3 keVee) events in a p-type point contact Germanium detector \cite{Aalseth:2010vx}, the CoGeNT Collaboration recently released the results of fifteen months of cumulative data-taking. These data provide evidence for the presence of a modulated component, with a statistical significance of 2.8$\sigma$, of unknown origin, compatible in principle with a light-mass WIMP \cite{Aalseth:2011wp}, featuring a mass around or slightly below 10 GeV, and a spin-independent scattering cross section with nucleons on the order of $10^{-40}\ {\rm cm}^{-2}$. The WIMP interpretation is in tension with results from the CDMS-II experiment, that recently carried out a dedicated low recoil-energy threshold analysis \cite{Ahmed:2010wy}, and with results from the XENON10 \cite{Angle:2007uj, Angle:2011th} and XENON100 \cite{Aprile:2010um, Aprile:2011hi} experiments. Ref.~\cite{Collar:2011wq}, however, critically argues that the limits presented in \cite{Angle:2011th} and in \cite{Aprile:2011hi} are, for various and different reasons, overstated, and that the parameters space compatible with a WIMP interpretation of the CoGeNT signal is presently not ruled out by either result. Notice that this claim is at present highly controversial - Ref.~\cite{Sorensen:2011bd} and \cite{Bezrukov:2010qa}, for example, evaluate and essentially refute the possibility of a low $L_{\rm eff}$ envisioned in Ref.~\cite{Collar:2011wq} as a caveat to the XENON results ruling out the CoGeNT signal. In addition, the note \cite{Collar:2011kf} argues that due to entirely different background cuts for the two detectors, the CDMS-II and CoGeNT measured irreducible spectra might actually be in agreement, and compatible with a $\sim7$ GeV WIMP signal. Remarkably, the CoGeNT signal is also in principle compatible with the annual modulation observed by DAMA/LIBRA if interpreted as WIMP recoils \cite{Bernabei:2008yi} (see also \cite{Hooper:2008cf}). Clearly, further experimental information is in order to settle the question, including closer inspection of detector issues, and new data from other experiments, such as CRESST, that also, tentatively, presented tantalizing signal excesses \cite{cresst}.

%interestingly, a Fermi excess has been claimed from the GC - warning, important effect of diffuse background, unresolved sources etc describe models that could fit the signal - cite buckley tait et al
Interestingly, an excess gamma-ray emission from the general direction of the Galactic Center region has been claimed recently to be present in the first two years of data collected by the Fermi-Large Area Telescope (LAT) \cite{Hooper:2010mq}. Under the assumptions of Ref.~\cite{Hooper:2010mq} for the diffuse and point-source background, the morphology and spectrum of the emission is in principle compatible with a 7-10 GeV mass WIMP annihilating dominantly into a pair of tau leptons, with a thermally averaged cross section times relative velocity in the range  $\langle\sigma v\rangle\sim5\times 10^{-27}$ to $5\times 10^{-26}\ {\rm cm}^3/$s. Ref.~\cite{Ellis:2011du} also recently compared the Fermi LAT data with predictions within the constrained minimal supersymmetric extension of the Standard Model.

%also intriguing is possibility to relate the haze to this
Ref.~\cite{Hooper:2010im} also recently entertained the intriguing possibility that the origin of the claimed gamma-ray excess from the center of the Galaxy might be related (via the produced relativistic electrons and positrons synchrotron-radiating at radio frequencies in the magnetic fields of the Galaxy) to an explanation for the intensity, spectrum and morphology of the so-called WMAP haze -- the unexplained excess at radio frequencies correlating with the innermost few degrees from the Galactic Center \cite{Dobler:2007wv}. 

It must be noted that data from the Galactic Center region appear, from the analyses and arguments presented in Ref.~\cite{2011ApJ...726...60C, Crocker:2010qn} to be consistent with a purely astrophysical origin, possibly connected with physics in the region surrounding the central supermassive black hole Sag A*. In particular,  Fermi LAT gamma-ray data from the Galactic Center region was found in \cite{Boyarsky:2010dr} to be entirely compatible with a standard background model consisting of a diffuse component plus point sources with different spectra from what was assumed in Ref.~\cite{Hooper:2010mq}.
 As is the case for the direct detection results, a more in-depth and detailed analysis and modeling of the gamma-ray data in the complex and crowded region of the Galactic Center seems timely and in order.

Given the tantalizing claims for WIMP signals, it is by all means tempting to accommodate those signals within a coherent particle-physics framework for particle dark matter. A model-independent analysis of particle physics scenarios that could accommodate the results reviewed above was recently given in Ref.~\cite{Buckley:2010ve}.  A specific, minimal realization of a beyond-the Standard Model dark matter framework that also could closely match the claimed direct and indirect detection excesses was outlined in Ref.~\cite{Logan:2010nw} (for a closely related supersymmetric model see also Ref.~\cite{Marshall:2011mm}). 

Here, we are concerned with a closer and detailed inspection of the ``lepton-specific'' two-Higgs-doublet model with singlet scalar dark matter proposed in Ref.~\cite{Logan:2010nw}. Following \cite{Logan:2010nw}, we will hereafter indicate this scenario with the acronym 2HDM-L. This class of models was, in a broader context, previously considered in Ref.~\cite{Goh:2009wg}, with an emphasis on larger masses and on a dark matter interpretation of the charged leptonic cosmic ray anomalies reported by PAMELA and ATIC. Early work on the phenomenological implications of general two-Higgs doublet models, including a classification of the relevant Yukawa interaction structures, was given in Ref.~\cite{Barger:1989fj}. More recently, the phenomenology of 2HDM with general Yukawa structures have been addressed in several studies, including Ref.~\cite{Aoki:2009ha, Su:2009fz, Logan:2009uf}. A recent study that addressed the dark matter phenomenology of a class of models closely related to the one under investigation here was given in Ref.~\cite{Aoki:2009pf}. The 2HDM-L plus singlet structure can also lead to phenomenologically viable baryogenesis and neutrino mass generation models, as discussed in Ref.~\cite{everything}. 

As we review in the following section, the model consists of a scalar singlet plus two SU(2) doublet scalar fields, one coupling to the quark sector and one coupling to the lepton sector. Ref.~\cite{Logan:2010nw} uses approximate expressions for the pair-annihilation cross section and for the relic density, and argues that matching the claimed gamma-ray excess and the observed dark matter relic density via thermal production might be possible. Again resorting to approximate expressions, Ref.~\cite{Logan:2010nw} argues that the direct detection cross section for the 2HDM-L dark matter model with a dark matter particle around 7-10 GeV predicts a spin-independent dark matter-nucleon cross section which is one order of magnitude below the required cross section to explain the CoGeNT anomalous signal. 

Ref.~\cite{Logan:2010nw} studies in some detail the collider signatures of the 2HDM-L model, showing that the Standard-Model-like Higgs generically acquires a large invisible width to the dark matter scalar, while the lepton-friendly Higgs decays predominantly invisibly with a few percent branching ratio into $\tau^+\tau^-$. While discovery of the predicted (heavier) Higgs structure is challenging at the LHC, Ref.~\cite{Logan:2010nw} argues that it would be straightforward at an $e^+e^-$ linear collider with a 500 GeV center of mass energy.

In the present study, we carry out a comprehensive study of the 2HDM-L parameter space, including detailed  and complete expressions for the relevant cross sections and for the dark matter thermal relic abundance. Our main result is that we find several models with the right mass, spin-independent and pair-annihilation cross sections to explain the direct and indirect signals reviewed above, and with a thermal relic density matching the observed dark matter abundance in the universe. We therefore argue that the 2HDM-L setup is particularly compelling as  a framework that would give a comprehensive WIMP interpretation to recent experimental and observational claims.

The remainder of this study consists of a review of the 2HDM-L model including constraints from electroweak precision measurements and perturbativity (sec.~\ref{model}), details on the relevant direct and indirect cross sections (sec.~\ref{2hdml8}), our model parameter space scan results (sec.~\ref{scan}), and, finally, our discussion and conclusions (sec.~\ref{disc}).

%%%%%%%%%%%%%%%%%%%%%%%%%%%%%%%%%%%%%%%%%%%%%%%%%%%%%%%%%%%%%%%%%%%%%%%%%%%%%%%%%%%%%%%%%%%%%%%%%%%%%%%%%
%%%%%%%%%%%%%%%%%%%%%%%%%%%%%%%%%%%%%%%%%%%%%%%%%%%%%%%%%%%%%%%%%%%%%%%%%%%%%%%%%%%%%%%%%%%%%%%%%%%%%%%%%
\section{The Model}
\label{model}
The model considered in \cite{Logan:2010nw} contains two complex SU(2)-doublet scalar fields $H_L$ and $H_Q$ that, through a Yukawa mechanism,
generate the masses of the charged leptons and the quarks, respectively, and a gauge-singlet real scalar field $S$, whose corresponding mass eigen-state
is the dark matter candidate. A $Z_2$ symmetry is imposed on the SU(2) fields, under which only $H_L$ and $e_{Ri}$ (right-handed
leptons) are odd. A distinct unbroken global $Z_2$ symmetry under which $S$ is odd is also assumed to ensure the stability of the dark matter candidate.
The scalar potential of the model is given by :

\begin{eqnarray}
\label{potential}
V&=&\mu_1 H_Q^{\dagger} H_Q+\mu_2 H_L^{\dagger}H_L+\frac{m_3}{2} S^2+\lambda_Q S^2 H_Q^{\dagger} H_Q+ \lambda_L S^2 H_L^{\dagger} H_L \nonumber \\
&+&\lambda_s S^4+\lambda_1 (H_L^{\dagger} H_L)^2+\lambda_2 (H_Q^{\dagger} H_Q)^2+\lambda_3 (H_Q^{\dagger}H_Q)(H_L^{\dagger} H_L) \nonumber \\
&+& \frac{\lambda_4}{2} ( (H_Q^{\dagger}H_L)^2+ h.c.)+\lambda_5 ( H_Q^{\dagger}H_L\, H_L^{\dagger} H_Q).
\end{eqnarray}
After electroweak symmetry breaking and the minimization of the potential we can write:
\begin{equation}
\label{fields}
\begin{array}{cc}
 H_L=\left(
\begin{array}{c}
H^{\prime +}_1\\
(v_1+H^{\prime}_1+i A^{\prime}_1)/\sqrt{2}
\end{array}
\right),
& H_Q=\left(
\begin{array}{c}
H^{\prime +}_2\\
(v_2+H^{\prime}_2+i A^{\prime}_2)/\sqrt{2}
\end{array}
\right).
\end{array}
\end{equation}
\\
Assuming CP conservation, the couplings $\lambda_{1-5}$ and the vacuum expectation values $v_{1-2}$ are taken to be real.
In what follows, un-primed particles denote mass eigenstates.

%%%%%%%%%%%%%%%%%%%%%%%%%%%%%%%%%%%%%%%%%%%%%%%%%%%%%%

\subsection{The Mass spectrum}
%\label{masses}
 The minimization of the tree-level potential leads to the following mass spectrum:

\begin{eqnarray}
\label{masses}
\label{m1}M^2_{H_1}&=& v_1^2 \lambda_1 + v_2^2 \lambda_2 - \sqrt{L^2\, v_1^2 v_2^2 + (v_1^2 \lambda_1 - v_2^2 \lambda_2)^2}; \\
M^2_{H_2}&=& v_1^2 \lambda_1 + v_2^2 \lambda_2 + \sqrt{L^2\, v_1^2 v_2^2 + (v_1^2 \lambda_1 - v_2^2 \lambda_2)^2}; \\
M^2_{A_2}&=&-v^2 \lambda_4; \\
M^2_{H_2^\pm}&=&-\frac{1}{2} v^2 (\lambda_4 + \lambda_5); \\
\label{m5}M^2_S&=&\frac{1}{2}(2\, m_3 + v_1^2 \lambda_L + v_2^2 \lambda_Q),
\end{eqnarray}
where $v^2=v_1^2+v_2^2=(246\ {\rm GeV})^2$ and $L=\lambda_3+\lambda_4+\lambda_5$. The mixing in the neutral CP-even sector is defined as:
\begin{eqnarray}
\label{mixing}
M_{H_1}&=& \cos \alpha\, {H^{\prime}_1} + \sin \alpha\, {H^{\prime}_2} ; \nonumber \\
M_{H_2}&=& -\sin \alpha\, {H^{\prime}_1} + \cos \alpha\, {H^{\prime}_2}.
\end{eqnarray}
We parametrize the model with the following set of independent variables: 
\begin{equation}
\label{parameters}
M_{H_1},\ M_{H_2},\ M_{A_2},\ M_{H_2^\pm},\ M_S,\ \lambda_L,\ \lambda_Q,\ \tan \beta=v_2/v_1,\ \alpha.
\end{equation}
The additional singlet quartic parameter is set to $\lambda_s=0.1$ throughout our analysis, since the ensuing phenomenology is widely independent of it. Additionally, for the purpose of the phenomenological study we are pursuing,
the masses of the pseudo-scalar and charged states play an insignificant role and will
be fixed to $120\ {\rm GeV}$ for the benchmark study. We are thus left with $6$ free parameters. We refer the Reader to Ref.~\cite{Logan:2010nw} for additional details on the model, including Feynman rules for the couplings of the scalar sector of the theory to quarks, leptons and gauge bosons.

%%%%%%%%%%%%%%%%%%%%%%%%%%%%%%%%%%%%%%%%%%%%%%%%%%%%%%

\subsection{Constraints on the model}
\label{consts}
In addition to the requirement that the scalar potential be bounded from below, we imposed the following theoretical and experimental constraints on the model:

\begin{itemize}
%%% item 1
\item {\it Electroweak precision tests}\\ 
\noindent 
  The oblique parameters $S,T,U$ provide tight constraints on generic theories beyond the Standard Model~\cite{Peskin:1991sw}.
  For models with additional scalars only, the $T$ parameter is the most relevant one, as scalars offer negligible contributions 
  to $S$ and $U$\cite{Grimus:2008nb,Barbieri:2006dq}. The stringent bounds on the $T$ parameter from electroweak
  measurements \cite{Nakamura:2010zzi} are:
  $$-0.08\leq T \leq0.14.$$
  Following the notation of \cite{Grimus:2007if}, the $T$ oblique parameter for the 2HDML-L model is :

\begin{eqnarray}
\label{delta}
\Delta \rho &=&  \frac{g^2}{64 \pi^2 M_W^2}\left[ F(M_{H_2^+}, M_{A_2}) + 3(F(M_W, M_{H_1}) - F(M_Z, M_{H_1})) \right. \nonumber  \\ & &
\left. \cos(\alpha-\beta)^2(F(M_{H_2^+}, M_{H_2})+F(M_{A_2},M_{H_2})+3F(M_W, M_{H_1})-3F(M_Z, M_{H_1}) )\right. \nonumber  \\ & &
\left. \sin(\alpha-\beta)^2(F(M_{H_2^+}, M_{H_1})+F(M_{A_2},M_{H_1})+3F(M_W, M_{H_2})-3F(M_Z, M_{H_2})) \right] \nonumber \\& &
	= \alpha_{em} T,
\end{eqnarray}
where $\alpha_{em}$ is the fine-structure constant and the function $F$ is defined as ($x, y>0$):

\begin{equation}
\label{Eq:F}
F \left( x, y \right) \equiv
\left\{ \begin{array}{ll}
{\displaystyle \frac{x+y}{2} - \frac{xy}{x-y}\, \ln{\frac{x}{y}}}
&\Leftrightarrow\ x \neq y,
\\*[3mm]
0 &\Leftrightarrow\ x = y.
\end{array} \right.
\end{equation}

 %%% item 2
\item {\it Perturbativity}\\ 
\noindent 
  Following Ref.~\cite{Su:2009fz} we impose perturbativity bounds on all the couplings of the model:
\begin{equation}
\label{pert}
\lambda_i \lsim 7 \ll{4 \pi} \hspace{0.5cm}i=1,...,5\hspace{0.05cm}
\end{equation}
 This requirement leads to an upper bound on the masses of the higgses of $\sim850$ GeV.
The lower bound is conservatively taken to be $114$ GeV, in accord with LEP-II searches for the Standard Model Higgs \cite{lephiggs}.
The upper bound on the masses of the charged and pseudo-scalar states is $\sim630$ GeV. For a discussion of parameter space bounds from perturbative unitarity, see also Ref.~\cite{pertunit}.

\item {\it Flavor Physics}\\ 
\noindent Extremely significant constraints on this model come from flavor physics, and in particular from measurements of BR($b\to s\gamma$), which in the presence of a charged Higgs boson receives important new physics contributions. We used here the results of Ref.~\cite{Su:2009fz} to implement that constraint. We find that other flavor observables are less constraining for this model than BR($b\to s\gamma$).
\end{itemize}

%%%%%%%%%%%%%%%%%%%%%%%%%%%%%%%%%%%%%%%%%%%%%%%%%%%%%%%%%%%%%%%%%%%%%%%%%%%%%%%%%%%%%%%%%%%%%%%%%%%%%%%%%
%%%%%%%%%%%%%%%%%%%%%%%%%%%%%%%%%%%%%%%%%%%%%%%%%%%%%%%%%%%%%%%%%%%%%%%%%%%%%%%%%%%%%%%%%%%%%%%%%%%%%%%%%

\section{Dark Matter in the 2HDM-L Model}
\label{2hdml8}
As detailed in the introduction, recently Hooper and Goodenough \cite{Hooper:2010mq} argued that an excess gamma-ray emission might exist in the inner
region of the Galactic Center (within approximately $1.25^{\circ}$) and be due to a dark matter particle in the
mass range $7-10\ {\rm GeV}$ annihilating primarily to tau leptons. We start our analysis by focusing on the 
slice with $M_S=8\ {\rm GeV}$ of the parameter space.
We then randomly scan over the other free parameters in the model (Eq.~(\ref{parameters}) above) and apply
the constraints on thermal relic density, indirect detection cross-section and the latest direct detection 
spin-independent cross-section exclusion bounds. We make no approximations in the expressions for all relevant cross sections.

As regards dark matter detection channel other than direct detection under scrutiny here, it is worthwhile to mention neutrino telescopes, that in principle provide very stringent constraints to models such as those under investigation here. Super-Kamiokande data can provide rather tight bounds on models with the mass range and pair-annihilation final states we considered here, as recently shown in Ref.~\cite{Kappl:2011kz}. While some of the models considered here might be in tension with Super-Kamiokande data, it is difficult to discern it directly from e.g. Fig.~3 in Ref.~\cite{Kappl:2011kz}, given that for the present models one has an admixture of s- and p-wave annihilation, and a mixture of final states that have significantly different constraints from neutrino telescopes.

%%%%%%%%%%%%%%%%%%%%%%%%%%%%%%%%%%%%%%%%%%%%%%%%%%%
\begin{figure*}[!t]
\mbox{\includegraphics[width=0.65\textwidth,clip]{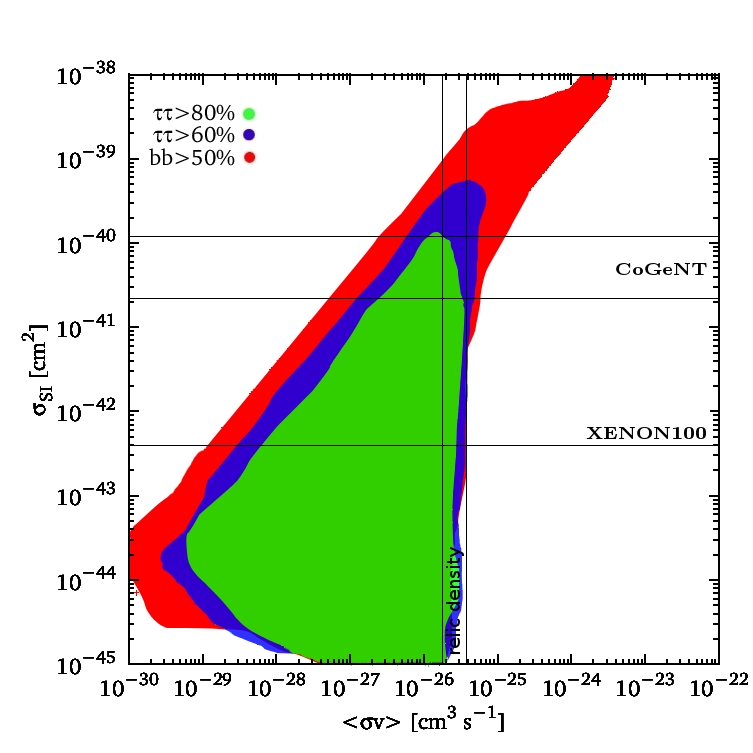}}
\caption{Spin-independent DM scattering cross section off protons versus annihilation cross-section
times the relative velocity of the DM particles.
Vertical lines guide the eye towards models with good cosmological relic density, whereas horizontal ones represent the
``CoGeNT region'' and the lower bound of XENON100. Regions where $\tau^{+}\tau^{-}$ final state accounts for
less than $50\%$ and more than $60\%$ and $80\%$ of the total annihilation channels are shown.
We fix $M_S=8{\rm GeV}$ and $M_{A_2}=M_{H_2^+}=120\ {\rm GeV}$.}\label{ddid1}
\end{figure*}
%%%%%%%%%%%%%%%%%%%%%%%%%%%%%%%%%%%%%%%%%%%%%%%%%%%

%%%%%%%%%%%%%%%%%%%%%%%%%%%%%%%%%%%%%%%%%%%%%%%%%%%%%%

%\subsection{Annihilation mechanism}
For a particle to be a viable cosmological dark matter candidate -- barring episodes of entropy injection (see e.g. \cite{Wainwright:2009mq}) -- the pair annihilation cross section must be large enough to prevent over-production in the early Universe via standard thermal freeze-out. For the particular
mass range we are interested in, and considering the Yukawa suppression for light fermions, the main kinematically open channels are annihilations to tau leptons
and to b quarks via $H_1$ and $H_2$ $s$-channels. Notice that there are no co-annihilation channels open in this model. 
%Also, due to the mixing (cf. eq.(\ref{mixing})) there is no simple dependence
%of the interaction vertices on $\lambda_L$ and $\lambda_Q$.

%%%%%%%%%%%%%%%%%%%%%%%%%%%%%%%%%%%%%%%%%%%%%%%%%%%%%%
%\subsection{Relic density and Direct Detection}
%\label{dd8}
In order to study the viable region of this benchmark we perform a random scan over the 6-dimensional
parameter space, fixing, as specified above, $M_{A_2}=M_{H_2^+}=120\ {\rm GeV}$. We compute the dark matter relic abundance
and the spin-independent scattering cross section off of protons with the {\tt micrOMEGAs} code
\cite{Belanger:2010gh,Belanger:2008sj}. Notice that to maximize the number of points passing the constraint on the
$T$ parameter we however tune, when necessary, $M_{H_2^+}$ to get a value in the correct range (in practice, we solve Eq.~(\ref{delta}) finding the value of $M_{H_2^+}$ for which $\Delta\rho=0$, for values of $M_{H_2^+}$ compatible with the LEPII constraints on the charged Higgs boson mass). Again, the tuning of this parameter does not affect quantities relevant for dark matter phenomenology.

In Fig.~\ref{ddid1} we show the result of a scan in the plane $(\sigma_{SI}$-$\langle\sigma v\rangle)$. Depending on the 
parameters of the model, the final states of dark matter annihilation (via $s$-channel $H_1$ and $H_2$ exchange) are mainly
tau leptons and b-quark pairs. Notice that in the absence of resonances or coannihilation, a thermal relic density corresponding to the observed dark matter density is found for models with a pair-annihilation cross section of a few $\times 10^{-26}\ {\rm cm}^3/$s. To guide the eye, we indicate the relevant range with vertical lines in the plot.

The fit to gamma-ray data obtained in Ref.~\cite{Hooper:2010mq} requires at least $60\%$ of the 
final states to be tau leptons. We explicitly show with different colors points where $\tau^{+}\tau^{-}$ final state accounts
for less than $50\%$ (red) and more than $60\%$ (blue) and $80\%$  (green) of the total annihilation channels.
Currently, the experiment that, at face value, puts the most stringent exclusion limits is XENON100 \cite{Aprile:2011hi} that
reaches a spin-independent cross-section of $\sim 4\times10^{-43}\mbox{cm}^2$ for a dark matter particle of
$8 {\rm GeV}$. It is clear from the plot that the 2HDM-L model has large parameter space regions compatible with cosmology
(relic density) and with the XENON100 null results, while featuring a dominant $\tau^+\tau^-$ annihilation final state. 

Interestingly enough, we find regions that are compatible with the annual modulation signal reported by the CoGeNT collaboration \cite{Aalseth:2011wp}, although imposing the
relic density constraint reduces significantly the region with a $\tau^{+}\tau^{-}$ annihilation final state dominating with more than $\sim60\%$ branching ratio. The fact that the b-quark-dominated region is a thin line (i.e. there exists a strong correlation between direct and indirect dark matter detection rates), whereas the tau-dominated regions are wider on the plane is due to the different contributions entering the two precesses of annihilation and the scattering cross-section off
protons.
Indeed, for the scattering off protons the parameter dependence of the vertices (except the mass of the quarks) is
essentially the same as for the annihilation into quarks, since it is basically the same process
in a $t$-channel. We have then that the ratio $\sigma_{SI}/\langle\sigma v\rangle$ is constant. This argument does not hold when the
annihilation proceeds through leptons because, in that case, the two SU(2) Higgses couple differently to leptons and quarks. 

An illustrative analytical derivation of
this result is as follows: The annihilation cross-sections into fermions can be approximated as \cite{Logan:2010nw}
\begin{equation}
  \label{sigmav}
  \langle\sigma v\rangle = \frac{N_c m_f^2}{4\pi}\left[1 - \frac{4 m_f^2}{s}\right]^{3/2}C_f^2,
\end{equation}
where $N_c$ is the number of colors of the species $f$, $s$ is the square of
the center-of-mass energy and $C_f$ includes all the contributions of the 6 parameters of the model and
are given by:
\begin{eqnarray}
  C_l &=& 
	  \frac{(\lambda_Q \tan\beta \cos\alpha - \lambda_L \sin \alpha)\sin\alpha}{s - M_{H_2}^2}
	  - \frac{(\lambda_Q \tan\beta \sin\alpha + \lambda_L \cos\alpha)\cos\alpha}{s - M_{H_1}^2},
  \nonumber \\
  C_q &=& 
	  -\frac{(\lambda_Q \cos\alpha - \lambda_L \cot\beta \sin\alpha)\cos\alpha}{s - M_{H_2}^2}
	  - \frac{(\lambda_Q \sin\alpha + \lambda_L \cot\beta \cos\alpha)\sin\alpha}{s - M_{H_1}^2}.
\label{eq:Cs}
\end{eqnarray}

The spin-independent cross-section of scattering off the proton is instead:
\begin{equation}
  \label{sigmaSI}
  \sigma_{SI} = \frac{m_p^4}{2 \pi (m_p + M_S)^2} C_q^2 F_N^2,
\end{equation}
where $m_p$ is the proton mass, $C_q$ is the same function that appears in Eq.~(\ref{sigmav}) evaluated at $s = 0$, and 
$F_N$ is a constant function of nuclear form factors, for more detail see e.g Ref.~\cite{Ellis:2000ds}.
Therefore, one has an approximate linear relation between $\sigma_{SI}$ and $\langle\sigma v\rangle$, the slope of which depends only on the mass
of the DM particle.
We will use this result in the next section where we perform a full scan including varying $M_S$.
Table \ref{points} presents sets of parameters that pass the constraints
listed in \ref{consts}, have the cosmological relic density and annihilate predominately to tau leptons.
Models A and B are compatible with the most stringent bounds from XENON100, whereas C and D are compatible with CoGeNT for
different values of $\tan \beta$ and low-mass Higgs.

\begin{table}[t!]
\begin{center}
\begin{tabular}{|c|c|c|c|c|c|c|c|c|c|c|}
\hline
Model & $M_{H_1}$ & $M_{H_2}$ & $\lambda_L$ & $\lambda_Q$ & $\sin \alpha$ & $\tan\beta$ & $\sigma_{SI} [\mbox{cm}^2]$ & $\Omega h^2$ & $\langle\sigma v\rangle$ [$10^{-26}\ {\rm cm}^3/{\rm s}$] ]\\
\hline
A& 114.8 & 177.1& 6.75 & -0.55 & 0.638 & 5.67 & 6.79$\times10^{-45}$ & 0.12 & 2.22 \\
B& 114.7 & 270. & 6.87 &  -0.21 & 0.01 & 5.79 & 2.17$\times10^{-43}$ & 0.12 &  2.24 \\
C& 117. & 163.  & -3.9 & 1.01 & 0.169 & 7.1 & 8.69$\times10^{-41}$ & 0.12 & 2.14 \\
D& 114.6 & 162.3 & -0.48 & 0.87 & 0.25&  6.62 & 8.04$\times10^{-41}$ &  0.091 & 2.96\\
\hline
\end{tabular}
\caption{Examples of set of parameters for the $M_S=8\  {\rm GeV}$ benchmark}\label{points}
\end{center}
\end{table}

%%%%%%%%%%%%%%%%%%%%%%%%%%%%%%%%%%%%%%%%%%%%%%%%%%%%%%%%%%%%%%%%%%%%%%%%%%%%%%%%%%%%%%%%%%%%%%%%%%%%%%%%%
%%%%%%%%%%%%%%%%%%%%%%%%%%%%%%%%%%%%%%%%%%%%%%%%%%%%%%%%%%%%%%%%%%%%%%%%%%%%%%%%%%%%%%%%%%%%%%%%%%%%%%%%%

\section{Full parameter space scan}
\label{scan}
\begin{figure*}[!t]
\includegraphics[width=0.65\textwidth,clip]{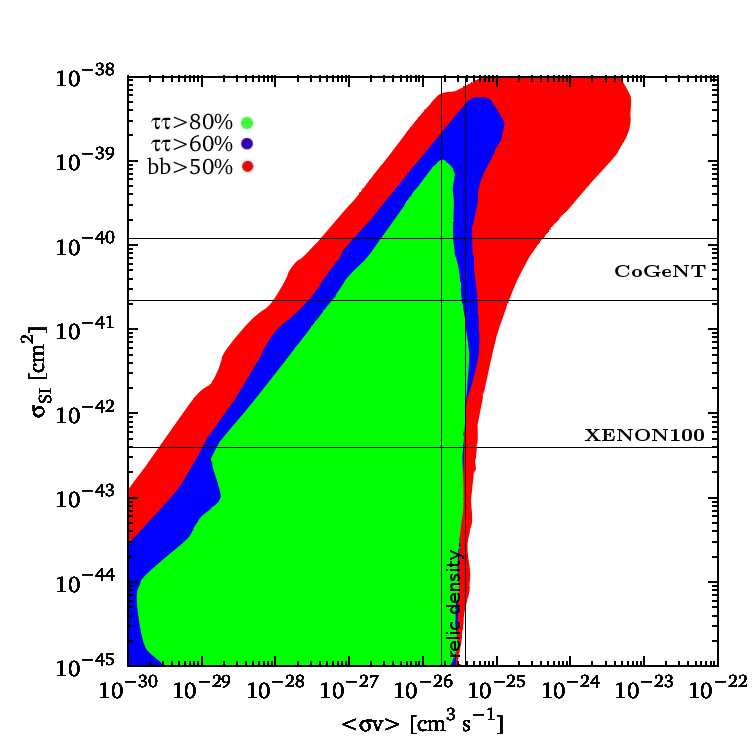}
\caption{Spin-independent dark matter scattering cross section off protons versus annihilation cross-section
times the relative velocity.
Vertical lines bound models with good cosmological relic density whereas horizontal ones represent the
CoGent region. Regions where $\tau^{+}\tau^{-}$ final state accounts for
less than $50\%$ and more than $60\%$ and $80\%$ of the total annihilation channels are shown.}\label{ddid2}
\end{figure*}

\begin{figure*}[!t]
\mbox{\includegraphics[width=0.7\textwidth,clip]{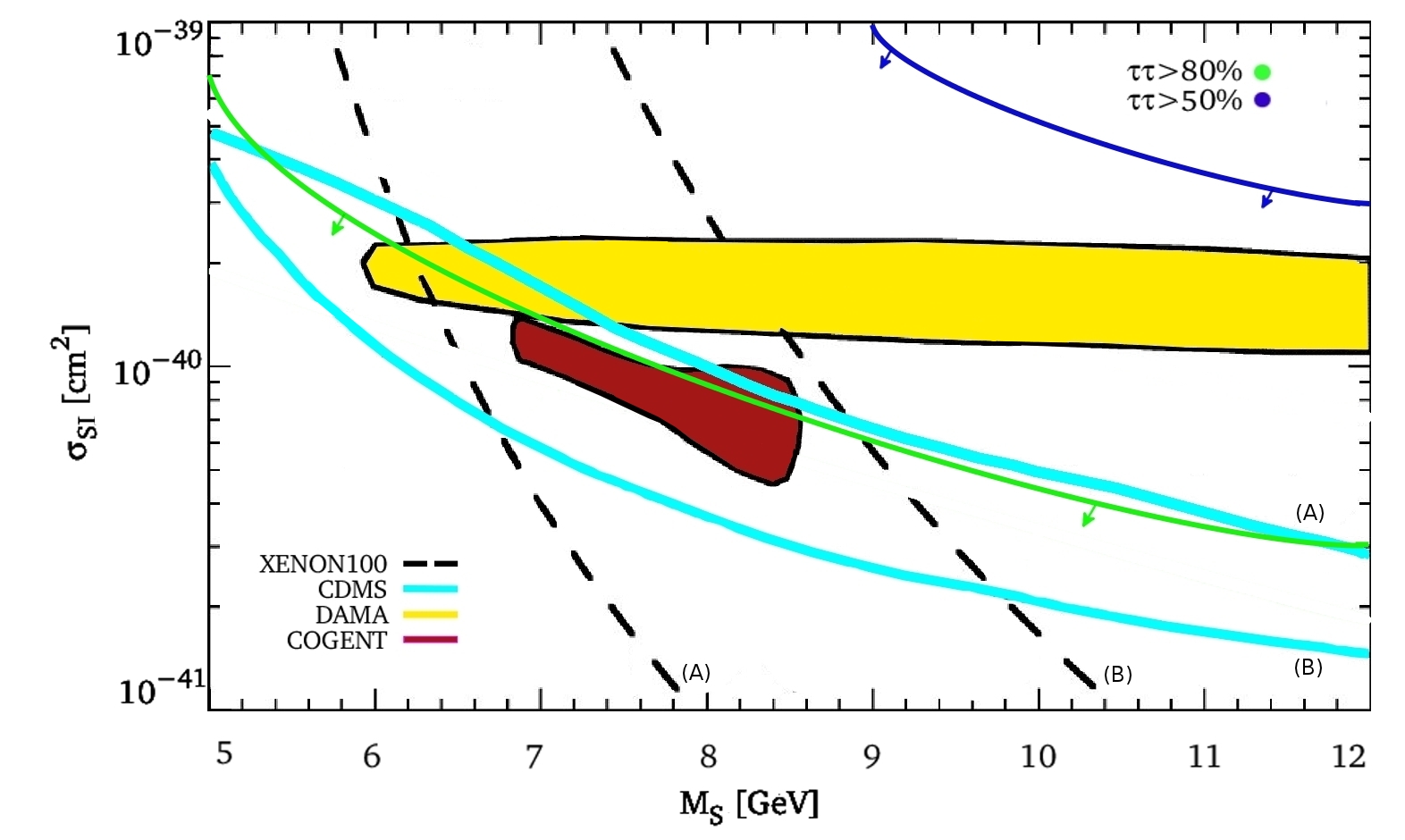}}
\caption{Spin-independent DM scattering cross-section off protons as a function of the mass of DM
with the regions of DAMA/LIBRA (without channeling)~\cite{Hooper:2010uy} and COGENT~\cite{Aalseth:2011wp} and the
sensitivity lines of CDMS and XENON100. For XENON100: (A) corresponds to the collaboration limits
~\cite{Aprile:2011hi} and (B) to a sensitivity obtained under different assumptions~\cite{Collar:2011wq}.
For CDMS: (A) is the CDMS limits~\cite{Akerib:2010pv} and (B) corresponds to CDMS-II results \cite{Ahmed:2010wy},
disputed in Ref.~\cite{Collar:2011wq}.
The blue lines delimit the region where lepton final states start dominating and the green one delimits
the region where the tau lepton final state dominates with more than $80\%$.}\label{dd}
\end{figure*}

We proceed now to a more general scan, to probe the low mass region of this model more systematically.
We follow the same strategy as for the study of the benchmark but we relax the constraints on the masses
$M_S$ and $M_{A_2}=M_{H_2^+}=120\ {\rm GeV}$. We vary linearly $M_S$ from $5\ {\rm GeV}$ to $15\ {\rm GeV}$, and we
sample randomly $10^7$ points in the following ranges:
\begin{itemize}
\item {\it Scalar Masses}: from 114 GeV up to the relevant pertubative limit ($\sim 600$ GeV).
\item {\it Charged and Pseudo-scalar particles}: from $100$ GeV up to their perturbative limit ($\sim 450$ GeV ).
\item {\it $\tan \beta$}: from 0 to 50.
\item {\it $\sin \alpha$}: from 0 to 1.
\end{itemize}
We apply the constraints described in sec.~\ref{consts} to every parameter space point. The result of the scan are shown on the $(\sigma_{\rm SI}, \langle\sigma v\rangle)$ plane in Fig.~\ref{ddid2} and on the ($M_S,\sigma_{\rm SI}$) plane in Fig.~\ref{dd}.

%%%%%%%%%%%%%%%%%%%%%%%%%%%%%%%%%%%%%%%%%%%%%%%%%%%
%%%%%%%%%%%%%%%%%%%%%%%%%%%%%%%%%%%%%%%%%%%%%%%%%%%

%%%%%%%%%%%%%%%%%%%%%%%%%%%%%%%%%%%%%%%%%%%%%%%%%%%
%%%%%%%%%%%%%%%%%%%%%%%%%%%%%%%%%%%%%%%%%%%%%%%%%%%

%In order to simplify the interpretation we draw iso-level lines corresponding to various values of the masses (5, 8 and 10 GeV) on the $(\sigma_{SI},\langle\sigma v\rangle)$  plane
%for the pure $b \bar b$ (in black) and less than $40\%$ $b \bar b$ final states using the linear relation 
%obtained in the previous section. Notice that these lines effectively correspond to upper-bounds on the regions of interest, since 
%$\sigma_{SI}$ depends only on $C_q$ for a given mass. Therefore it is straightforward to
%see that to explain CoGeNT with a sensible thermal relic abundance, the mass of the dark matter cannot exceed $\sim~10\ {\rm GeV}$ which is
%consistent with what was inferred from the study of the Galactic Center in Ref.~\cite{Hooper:2010mq}.

In Fig~\ref{dd} we focus on direct detection only, and show the experimental sensitivities of DAMA, CoGeNT, XENON100
and CDMS \cite{Aalseth:2011wp}. Notice that different analyses obtain different regions compatible, at some given confidence level, with the DAMA/LIBRA modulation signal. For example, the shape of the parameter space compatible with DAMA/LIBRA obtained in e.g. Ref.~\cite{savageetal} and \cite{Hooper:2010uy} differ in the allowed mass range, and this effect depends upon different assumption on the choice of the bins used, the quenching factors, the velocity distribution, and the definition of the confidence levels. Here, we choose to employ the results of Ref.~\cite{Hooper:2010uy}, which emphasize the possibility that the CoGeNT and DAMA signals might have a common origin, which is an especially appealing scenario for the model we investigate here. Regions below the green line contain 2HDM-L models that are consistent with having the correct thermal relic abundance, and featuring a pair-annihilation final state branching ratio larger than $80\%$ into the $\tau^{+}\tau^{-}$ final state. Models below the dark blue line, instead, have branching ratios larger than $50\%$. The figure illustrates that we find ample parameter space in the 2HDM-L framework to accommodate the signals reported in direct dark matter detection with dark matter models with the right relic abundance that feature large branching ratios into $\tau^{+}\tau^{-}$.

%%%%%%%%%%%%%%%%%%%%%%%%%%%%%%%%%%%%%%%%%%%%%%%%%%%%%%%%%%%%%%%%%%%%%%%%%%%%%%%%%%%%%%%%%%%%%%%%%%%%%%%%%
%%%%%%%%%%%%%%%%%%%%%%%%%%%%%%%%%%%%%%%%%%%%%%%%%%%%%%%%%%%%%%%%%%%%%%%%%%%%%%%%%%%%%%%%%%%%%%%%%%%%%%%%%

\section{Discussion and Conclusions}
\label{disc}

We investigated the phenomenology of the dark matter candidate arising from a lepton-specific two-Higgs doublet model with a gauge-singlet scalar. We imposed on the parameter space both electroweak constraints and limits from LEP searches for the Higgs. We then calculated the spin-independent dark matter scattering cross section off of nucleons, and the pair-annihilation cross section. The two quantities have a very simple correlation for dominant b-quark annihilation final state, while the correlation is lost for tau-lepton final state. Using the {\tt micrOMEGAs} code we calculated the thermal dark matter relic abundance. 

We reached here somewhat different, and more optimistic conclusions than what reported on a recent study on the same model \cite{Logan:2010nw}. Specifically, we showed here that within the model under consideration one can account for the anomalous signals reported by CoGeNT, DAMA/LIBRA and the claims of a gamma-ray excess from the Galactic Center, while, at the same time, having a thermal dark matter relic abundance compatible with the observed universal dark matter density. Should the various anomalous signals discussed here pass forthcoming scrutiny, the simple dark matter framework we investigated in this study might indeed deserve additional attention.

\begin{acknowledgments}
\noindent  SP is partly supported by an Outstanding Junior Investigator Award from the US Department of Energy and by Contract DE-FG02-04ER41268, and by NSF Grant PHY-0757911. The Work of MSB was supported by the Spanish grants FPA2008-00319 and MULTIDARK Consolider CSD2009-00064 (MICINN), and Prometeo/2009/091 (Generalitat Valenciana), as well as by the EU network UNILHC PITN-GA-2009-237920 and an FPI fellowship from MICINN. MSB would like to thank the Santa Cruz Institute for Particle Physics at UCSC for kind hospitality.
\end{acknowledgments}

%\bibliography{boucenna_profumo_07.bib}

\end{document}